\documentclass[prb, preprint,showpacs,preprintnumbers,superscriptaddress, amsmath,amssymb]{revtex4}
\usepackage{graphicx}% Include figure files
\usepackage{dcolumn}% Align table columns on decimal point
\usepackage{bm}% bold math

\begin{document}

\preprint{APS/123-QED}

\title{Novel Magnetic and Thermodynamic Properties of Thiospinel 
  Compound CuCrZrS$_{4}$}% Force line breaks with \\

\author{Masakazu Ito}
%\email{showa@hiroshima-u.ac.jp}
\altaffiliation[Also at ]{Institute of Pure and Applied Physical Sciences, University
of California at San Diego, La Jolla, California 92093}%Lines break automatically or can be forced 
\affiliation{Department of Quantum Matter, ADSM, Hiroshima University, Higashi-Hiroshima 739-8530 Japan\\}
\author{Hiroki Yamamoto}
\affiliation{Department of Materials Science and Engineering, Muroran Institute of Technology, 27-1 Mizumoto-cho, Muroran 050-8585, Japan\\}
\author{Shoichi Nagata}
\affiliation{Department of Materials Science and Engineering, Muroran Institute of Technology, 27-1 Mizumoto-cho, Muroran 050-8585, Japan\\}
\author{Takashi Suzuki}%
\affiliation{Department of Quantum Matter, ADSM, Hiroshima University, Higashi-Hiroshima 739-8530 Japan\\}
%\email{tsuzuki@hiroshima-u.ac.jp}
\date{\today}% It is always \today, today,
             %  but any date may be explicitly specified

\begin{abstract}
 We have carried out dc magnetic susceptibility, magnetization and specific heat measurements on thiospinel CuCrZrS$_{4}$.
 Below $T_{\rm C}^{*} =$ 58 K, dc magnetic susceptibility and magnetization data show ferromagnetic behavior with a small spontaneous magnetization 0.27  $\mu _{\rm B}/$f. u..
In dc magnetic susceptibility, large and weak irreversibilities are observed below $T_{\rm f} =$ 6 K and in the range $T_{\rm f}< T < T_{\rm C}^{*}$ respectively. 
We found that there is no anomaly  as a peak or step in the specific heat at $T_{\rm C}^{*}$.   
\end{abstract}

% PACS, the Physics and Astronomy
                             % Classification Scheme.
%\keywords{Suggested keywords}%Use showkeys class option if keyword
                              %display desired
\maketitle

%\section{\label{sec:level1}First-level heading:\protect\\ The line
%break was forced \lowercase{via} \textbackslash\textbackslash}
\section{Introduction}
Chalcogenide spinels with the generic chemical formula $AB_2X_4$ are well known to show a number of interesting physical properties. For example, CuRh$_2$S$_4$\cite{Hagino, Maaren, Shelton, Bitoh} and CuRh$_2$Se$_4$\cite{Hagino, Maaren, Shelton, Shirane} are superconductors,  CuIr$_2$S$_4$ shows a temperature-induced metal-insulator transition due to a new type charge ordering \cite{Nagata, Furubayashi1, Radaelli} and FeCr$_{2x}$In$_{2(1-x)}$S$_{4}$ shows a reentrant spin glass behavior from a long range antiferromagnetic order.\cite{Moreno}
Recently, we found the amazing phenomenon of a  pressure-induced superconductor-insulator transition in CuRh$_2$S$_4$.\cite{Ito1, Ito2}
 CuCr$_2$S$_4$, the parent compound to CuCrZrS$_{4}$, is a metallic compound which shows a ferromagnetic transition at a Curie temperature $T_{\rm C} =$ 377 K with a saturation magnetization of 5 $\mu_{\rm B}$/f. u..\cite{Lotgering1, Kanomata, Obayashi, Robbins}
 The $B$ site Cr ions in CuCr$_2$S$_4$ have valences Cr$^{3+}$ (spin angular momentum $S$ $=$ 3/2, magnetic moment $m$ $=$ 3 $\mu_{\rm B}$) and Cr$^{4+}$ ($S$ $=$ 1, $m$ $=$ 2 $\mu_{\rm B}$). \cite{Lotgering1, Lotgering2, Robbins, Krok}
This mixed valence state of Cr ions results in the metallic conductivity of CuCr$_2$S$_4$. 
 Ferromagnetism with 5 $\mu_{\rm B}$/f. u. in this compound is attribute to the double-exchange interaction between the Cr$^{3+}$ and Cr$^{4+}$ anions via the conduction electrons.\cite{Zener1, Stapele, Kimura} 

High-purity specimens of  CuCrZrS$_{4}$ were recently synthesized and investigated through transport and magnetic measurements by Iijima $et\ al.$.\cite{Iijima}  
When Zr ions occupy half of the $B$ sites of CuCr$_{2}$S$_{4}$, the temperature dependence of the electric resistivity changes from metallic to semiconducting with an energy gap of 79 K. 
 The Curie temperature simultaneously decreases to 60 K. 
 In addition, spin-glass behavior is observed below the freezing temperature $T_{\rm f} =$ 10 K, namely, CuCr$_{2}$S$_{4}$ shows reentrant spin glass freezing from the long range ferromagnetic order.  
 It has been confirmed that the ionic configuration of this compound is Cu$^{1+}$Cr$^{3+}$Zr$^{4+}$S$^{2-}_{4}$. %hen
 In this letter, we report the dc magnetic susceptibility, magnetization and specific heat of  CuCrZrS$_{4}$. These results suggest that the ferromagnetic behavior of CuCrZrS$_{4}$ is not caused by the long-range ferromagnetic order.  
%
%
%\par
\section{Experimental}
 A polycrystalline specimen was prepared by a direct solid-state reaction as described  previously.\cite{Iijima}  High purity fine powders of Cu (99.99\%), Cr (99.99\%), Zr (99.9\%) and S (99.999\%) were mixed in stoichiometric ratio and were reacted in a quartz tube at 1023 K for 7 days.
A powder specimen was reground, pressed into a rectangular bar and sintered at 1023 K for 2 days.
 The magnetic measurements were carried out using a Quantum Design MPMS SQUID magnetometer.
  The dc magnetic susceptibility was measured as a function of temperature in 2 $\le \ T\ \le $ 300 K during warm up after cooling to 2 K both in the zero field (ZFC) and in the measuring field of $H\ =$ 100 Oe (FC). 
 The magnetization in the magnetic field range between $-45$ and $45$ kOe at 2 K was measured after ZFC.
 Specific heat measurements in the temperature range between 1.8 and 150 K were carried out using a Quantum Design PPMS which is operated by the thermal relaxation method.%
\section{Results}
\subsection{Magnetic properties}
%
%\par
Figure 1(a) shows the temperature $T$ dependence of dc magnetic susceptibility $\chi$ and inverse susceptibility $\chi^{-1}$ of  CuCrZrS$_{4}$ at 100 Oe in the temperature range between 2 and 300 K.  
 With decreasing $T$, $\chi$ increases rapidly below $T_{\rm C}^{*}\ =$ 58 K as seen at the Curie temperatures for ordinary ferromagnetic compounds.
 $T_{\rm C}^{*}$ is defined by the onset of uprise in $\chi$.  
 As shown by the solid line in Fig. 1(a), magnetic susceptibility above 75 K can be fitted by the Curie-Weiss law, $\chi $ $=$ $C/( T - \theta  ),$ with a paramagnetic Curie temperature $\theta$ $=$ 57 K and a Curie constant $C$ $=$ 1.534 emu/mol$\cdot $K. 
 The positive $\theta$ means the magnetic interaction in the system is ferromagnetic. 
 The effective magnetic moment $\mu _{\rm eff}$ obtained from $C$ is 3.48 $\mu_{\rm B}$, which is close to the value of 3.87 $\mu_{\rm B}$ expected for free Cr$^{3+}$ ion.
 As was pointed out by Iijima $et\ al.$,\cite{Iijima} a large irreversibility between $\chi_{_{\rm FC}}$ and $\chi_{_{\rm ZFC}}$ is observed in the low temperature range as shown in Fig. 1(b).
 This behavior is characteristic of spin-glass freezing. 
\begin{figure}
\begin{center}
%\figureheight{7cm}
\includegraphics[height=110mm,clip]{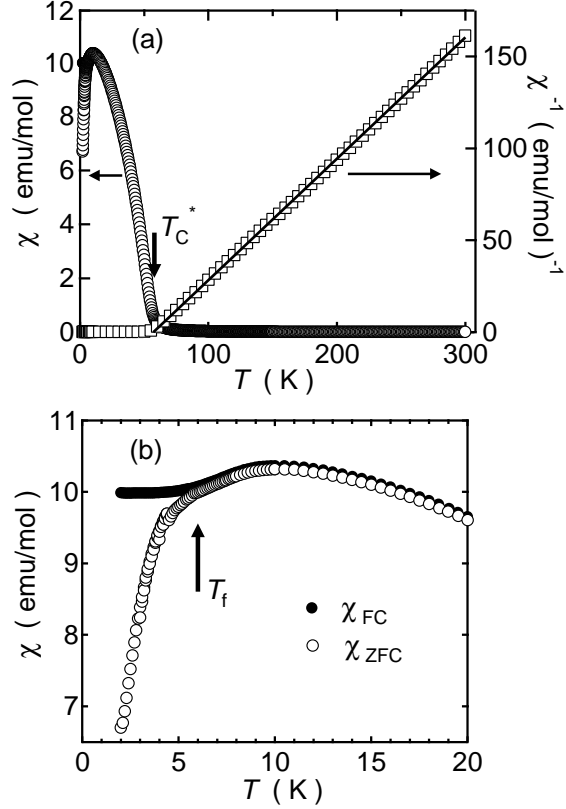}
\caption{(a) Temperature dependence of the dc magnetic susceptibility $\chi $ and inverse susceptibility $\chi ^{-1}$ under magnetic field $H =$ 100 Oe in the temperature range 2 $\le  T \le$ 300 K. The solid line is the best fit by Curie-Weiss law.
 (b) An expanded plot of $\chi $ between 2 and 20 K. The closed and open circles represent $\chi $ measured with field cooled (FC) and zero field cooled (ZFC), respectively. }
\label{fig:1}
\end{center}
\end{figure}
\par
 In order to discuss the irreversibility in more detail, the difference magnetic susceptibility $\Delta \chi $ $(\ = \chi_{_{\rm FC}}-\chi_{_{\rm ZFC}}\ )$ is calculated and shown in Fig. 2.
 $\Delta \chi $ shows an abrupt increase below $T_{\rm f}$ = 6 K with decreasing $T$.  
% A more important point is that
 $\Delta \chi $ has a non-zero value even in the range $T_{\rm f} \le T \le T_{\rm C}^{*}$ as shown in the inset of Fig.2. 

 Figure 3(a) shows the magnetization $M$ vs. field $H$ curve of CuCrZrS$_{4}$ at 2 K  in the range  $-45 \le H \le 45$ KOe.  
$M$ is not saturated in field up to 45 kOe. 
 In lower $H$ range, a small hysteresis loop is observed as shown in Fig. 3(b).
 The loop has a small coercive force of $H \sim$ 50 Oe which is defined as the field necessary to restore zero magnetization.  
 The value of spontaneous magnetization $M_{\rm 0}$, estimated by extrapolation to $H = 0$ Oe from the range $10 \le H \le 50$ kOe, is 0.27 $\mu_{\rm B}$/f. u., is close to the value reported by Iijima.\cite{Iijima} 
 This value is only $\sim$10\% of the value expected for a Cr$^{3+}$ ion. 
\begin{figure}
\begin{center}
%\figureheight{7cm}
\includegraphics[height=65mm,clip]{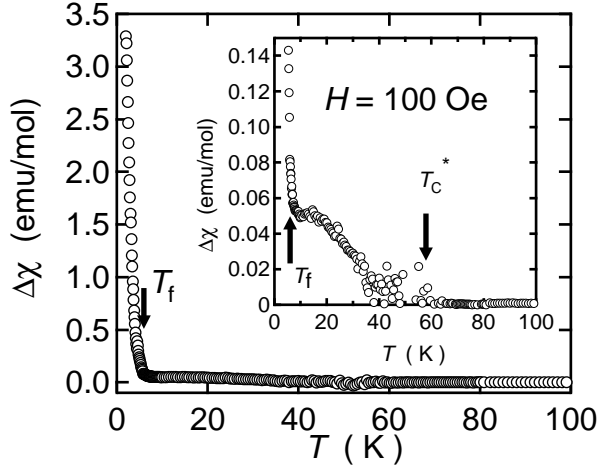}
\caption{Difference magnetic susceptibility $\Delta \chi $ as a function of $T$ in the range 2 $\le T \le$ 100 K.
The inset is a plot of $\Delta \chi $ with expanded vertical scale.}
\label{fig:2}
\end{center}
\end{figure}
\begin{figure}
\begin{center}
\includegraphics[height=130mm,clip]{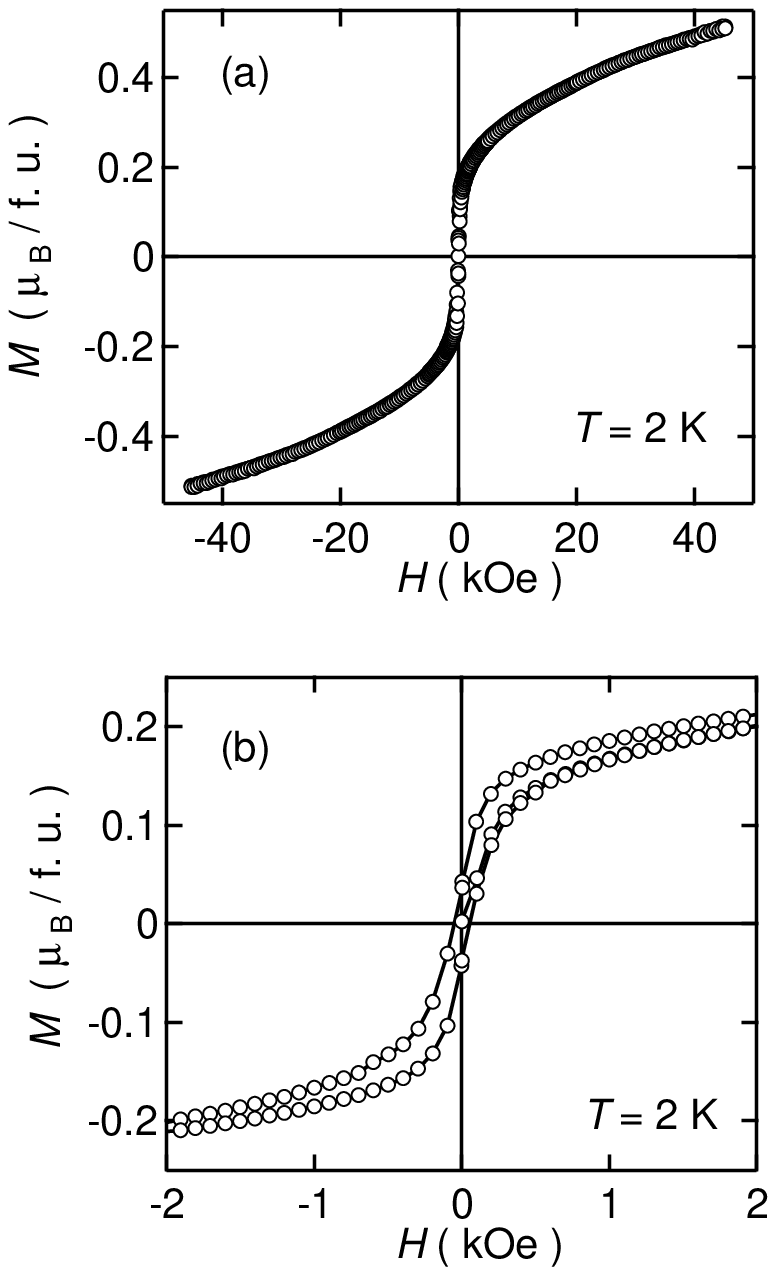}
\caption{Magnetization as a function of the magnetic field $H$ at 2 K.  (a) $M(H)$ in the range $-45 \le H \le 45$ kOe. (b) An expanded plot of $M(H)$ in the range $-2 \le H \le 2$ kOe.}
\label{fig:3}
\end{center}
\end{figure}
\subsection{Thermodynamic properties}
%
%\par
Figure 4(a) shows the temperature dependence of specific heat $C_{P}(T)$ of CuCrZrS$_{4}$ in the range 1.8 $\le T\le $ 400 K.
 At 300 K, $C_{P}(T)$ reaches 85 \% of that estimated from the Dulong and Petit law ($ \sim$ 175 J/K$\cdot$ mol). 
 Remarkably, neither a peak or a step is observed in $C_{P}$ at $T_{\rm C}^{*}$ as shown in Fig. 4 (b), despite that clear ferromagnetic  behavior is observed from magnetization. 
 Although specific heat measurements with a relaxation method are not suitable for obtaining the absolute value of $C_{P}$ which has a large temperature dependence around the 1st order transition, we can confirm whether there is an anomaly in $C_{P}$.
 The absence of an anomaly in the specific heat at $T_{C}^{*}$ is still unclear.
This unusual behavior will be discussed later. 
%  even at $T_{\rm C}^{*}$.
Figure 4 (c) is the plot of $C_{P}/T$ as a function of $T$.
A broad hump appears around $T_{\rm f}$. 
 It is well known that the specific heat of spin-glass systems such as (Eu, Sr)S\cite{Meschede} and AuFe\cite{Mirza} do not show a discontinuity as seen in the 1st and 2nd order transition but rather a broad hump around the spin-glass freezing temperature. %hen
 Our result suggests that spin-glass freezing occurs below $T_{\rm f}$. 
\begin{figure} 
\begin{center}
%\figureheight{7cm}
\includegraphics[height=195mm,clip]{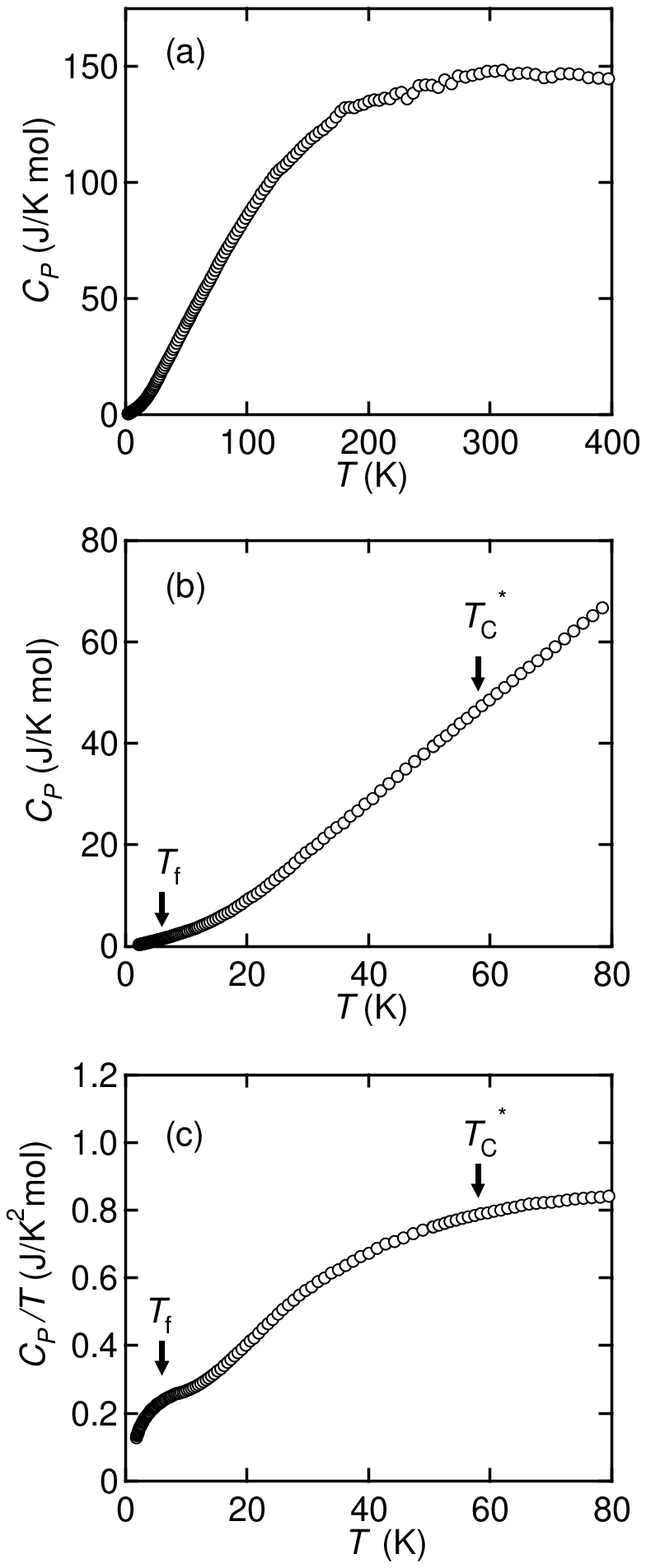}
\caption{(a) Temperature dependence of specific heat  $C_{P}$ in the range 1.8 $\le T \le$ 400 K. 
(b) An expanded plot of $C_{P}(T)$ in the range 1.8 $\le T \le$ 80 K.
(c) Temperature dependence of specific heat divided temperature, $C_{P}/T$ in the range 1.8 $\le T \le$ 80 K. 
}
\label{fig:4}
\end{center}
\end{figure}
\section{Discussion}
\subsection{Spin-glass behavior in low temperature}
  In general, spin-glass freezing originates from the frustration of the magnetic interaction in the system. 
  We will discuss the frustration in  CuCrZrS$_{4}$.
As mentioned above, CuCr$_{2}$S$_{4}$ has metallic conductivity which is attributed to the mixed valence of Cr ions (Cr$^{3+}$ and Cr$^{4+}$) and the ferromagnetism due to the double-exchange interaction between Cr ions via conduction electrons. 
 When half of the Cr ions of CuCr$_{2}$S$_{4}$ are substituted by Zr ions, the electric configuration is Cu$^{1+}$Cr$^{3+}$Zr$^{4+}$S$^{2-}_{4}$, and the transport properties become semiconductive.\cite{Iijima} This behavior can be explained by a variable range hopping (VRH) conduction.\cite{Furubayashi2} 
This means that the number of conduction electrons in CuCrZrS$_{4}$ decrease with decreasing temperature.
The decrease in the number of conduction electrons gives rise to the reduction of the double-exchange interaction and relative enhancement of the superexchange antiferromagnetic interaction between Cr$^{3+}$ anions via S$^{2-}$ cations at low temperature.  
The Cr$^{3+}$ anions in  CuCrZrS$_{4}$ have a tetrahedral network with each Cr$^{3+}$ anions. 
 When the interaction between Cr$^{3+}$ anions is antiferromagnetic, the tetrahedral network causes strong geometrical frustration in the system.
  This is similar to behavior in the pyrochlore crystal structure.\cite{Steven}
\subsection{Specific heat}
Now we discuss why there is no anomaly at $T_{\rm C}^{*}$ in $C_{P}(T)$ from the viewpoint of the phase with and without the ferromagnetic long range order below $T_{\rm C}^{*}$.  
\par
In the first case, one can consider the coexistence of the itinerant and the localized spins in $t_{2g}$ orbital.
 Some of Cr compounds have large density of states (DOS) which predominantly consist of one of the $t_{2g}$ in 3$d$ orbital of the Cr ions close to below the Fermi level that is occupied by the localized spins.\cite{Korotin, Raju}    
On the other hand, other $t_{2g}$ orbitals are strongly hybridized with $p$ orbitals of the anions in the compound and form the dispersive bands  which have the itinerant spins. This picture leads to the coexistence of the localized and the itinerant spins in the $t_{2g}$ orbital electrons.
The itinerant spins that belong to the $d$ ($t_{2g}$)-$p$ hybridized bands tend to polarize the localized spins because of Hund's rule coupling, resulting in a ferromagnetic exchange interaction which is often called a self-doped double exchange interaction.\cite{Raju}
This mechanism is very similar to that of the double exchange interaction between the localized $t_{2g}$ electrons and the itinerant electrons in the $d$ ($e_{g}$)-$p$ hybridized bands in the Zener model.\cite{Zener2}
Recently, Granado et al. studied the neutron scattering and specific heat of LaCrSb$_{3}$.\cite{Granado}
They reported that LaCrSb$_{3}$ shows the ferromagnetic long range order at $T_{\rm C}$ = 126 K and nevertheless no peak or step appears in specific heat at $T_{\rm C}$.
They suggested that unconventional magnetic behavior results from that the localized spins which are strongly polarized even  above $T_{\rm C}$ by  the self-doped double exchange interaction. 
 This situation can be responsible for entropy transfer to higher temperature, and released entropy at $T_{\rm C}$ is small.
 Consequently, an anomaly in $C_{P}$ at $T_{\rm C}$ is too small to be observed.
Although the band structure of  CuCrZrS$_{4}$ is not clarified yet, 
 if the localized moments in CuCrZrS$_{4}$ are strongly polarized by the itinerant electrons due to a mechanism similar to that of ferromagnetism in LaCrSb$_{3}$, no anomaly at $T_{\rm C}^{*}$ in specific heat might be explained.
\par
 As the second case, we suggest the possibility of successive spin-glass freezing at $T_{\rm f}$ and  $T_{\rm C}^{*}$.
 Existence of small ($T_{\rm f} \le T \le T_{\rm C}^{*}$) and large ($T \le T_{\rm f}$) irreversibility in $\Delta \chi $ is similar to successive spin-glass freezing seen in the spin-glass material Ni$_{0.42}$Mn$_{0.58}$TiO$_{3}$.\cite{Kawano}  
As mentioned in section IV-A, this system has competing ferromagnetic and antiferromagnetic exchange interactions.
With decreasing $T$, the number of conduction electrons decrease, and the superexchange interaction gradually grows stronger.
This promotes variation of the strength of geometric frustration with decreasing $T$, and 
might be the origin of the successive spin-glass freezing.
If spin-glass freezing occurs below $T_{\rm C}^{*}$, no discontinuously at $T_{C}^{*}$ in specific heat would be observed. %hen  
Unfortunately, the broad hump around $T_{\rm C}^{*}$ in specific heat which is characteristic of spin-glass freezing, is not observed due to the  difficulty of separating the phonon contribution from total specific heat. 
Appearance of small $M_{\rm 0}$ in the low temperature range might originate from the spin glass freezing with the short range ferromagnetic correlation. 
 In order to clarify why no discontinuously at $T_{C}^{*}$ in specific heat is observed, detailed studies are needed.
%hen
% 
%
%
%
%
%
\section{Conclusion}
We have studied the magnetic and thermodynamic properties of the thiospinel compound  CuCrZrS$_{4}$ which has been reported to show long-range ferromagnetic order at 60K and spin-glass freezing at 10 K. 
 The dc magnetic susceptibility and magnetization curve show ferromagnetic behavior below $T_{\rm C}^{*} =$ 58 K with a small spontaneous magnetization 0.27 $\mu _{\rm B}/$f. u..
With decreasing temperature, the irreversibility in dc magnetic susceptibility appears at $T_{\rm C}^{*}$, and increases abruptly at $T_{\rm f} =$ 6 K.
 In specific heat, no magnetic discontinuity is observed in the temperature range we measured. %hen 
 We discussed the origin of no magnetic discontinuity in specific heat from the viewpoint of the phase with and without the ferromagnetic long range order below $T_{\rm C}^{*}$. 
\begin{acknowledgments}
This work was partially supported by Grant-in-Aids for COE Research (No. 13CE2002), the Scientific Research (No.16740205, No. 17340113) from the Ministry of Education, Culture, Sports, Science and Technology of Japan,  and aid funds from Energia, Inc (Hiroshima) and the Asahi Glass Foundation (Tokyo).
\end{acknowledgments}

\end{document}